\documentclass[]{emulateapj}

\slugcomment{}
\shorttitle{Optical--IR Properties of Faint Submm Sources}
\shortauthors{Hatsukade et al.}

\let\lsim=\la
\let\gsim=\ga
\begin{document}

\title{Optical--Infrared Properties of Faint 1.3~mm Sources detected with ALMA}

\author{
	Bunyo Hatsukade\altaffilmark{1*},
	Kouji Ohta\altaffilmark{2},
	Kiyoto Yabe\altaffilmark{1,3},
	Akifumi Seko\altaffilmark{2},
	Ryu Makiya\altaffilmark{4},
	and Masayuki Akiyama\altaffilmark{5},
}

\affil{
\altaffilmark{1}National Astronomical Observatory of Japan, 2-21-1 Osawa, Mitaka, Tokyo 181-8588, Japan \\
\altaffilmark{2}Department of Astronomy, Kyoto University, Kyoto 606-8502, Japan \\
\altaffilmark{3}Kavli Institute for the Physics and Mathematics of the Universe (WPI), The University of Tokyo, 5-1-5 Kashiwanoha, Kashiwa, Chiba 277-8583, Japan \\
\altaffilmark{4}Institute of Astronomy, University of Tokyo, 2-21-1 Osawa, Mitaka, Tokyo 181-0015, Japan \\
\altaffilmark{5}Astronomical Institute, Tohoku University, Aramaki, Aoba-ku, Sendai, Miyagi 980-8578, Japan \\
\altaffilmark{}{\it Received 2015 February 12; accepted 2015 July 31}
}

\email{bunyo.hatsukade@nao.ac.jp}
\altaffiltext{*}{NAOJ Fellow}

\begin{abstract}
We report optical-infrared (IR) properties of faint 1.3~mm sources ($S_{\rm 1.3 mm} = 0.2$--1.0~mJy) detected with the Atacama Large Millimeter/submillimeter Array (ALMA) in the Subaru/{\sl XMM-Newton} Deep Survey (SXDS) field. 
We searched for optical/IR counterparts of 8 ALMA-detected sources ($\ge$4.0$\sigma$, the sum of the probability of spurious source contamination is $\sim$1) in a $K$-band source catalog. 
Four ALMA sources have $K$-band counterpart candidates within a 0\farcs4 radius. 
Comparison between ALMA-detected and undetected $K$-band sources in the same observing fields shows that ALMA-detected sources tend to be brighter, more massive, and more actively forming stars. 
While many of the ALMA-identified submillimeter-bright galaxies (SMGs) in previous studies lie above the sequence of star-forming galaxies in stellar mass--star-formation rate plane, our ALMA sources are located in the sequence, suggesting that the ALMA-detected faint sources are more like `normal' star-forming galaxies rather than `classical' SMGs. 
We found a region where multiple ALMA sources and $K$-band sources reside in a narrow photometric redshift range ($z \sim 1.3$--1.6) within a radius of $5''$ (42~kpc if we assume $z = 1.45$). 
This is possibly a pre-merging system and we may be witnessing the early phase of formation of a massive elliptical galaxy.
\end{abstract}


\section{Introduction}
Since the fraction of dust-obscured star formation to the total star formation increases with redshift \citep[e.g.,][]{take05, lefl05}, observations at infrared (IR) to millimeter/submillimeter (mm/submm) wavelengths are essential to understand the cosmic star formation history and the galaxy evolution. 
Deep and wide-field surveys uncovered a new population of mm/submm--bright galaxies at high redshifts (SMGs) \citep[see][for a review]{blai02}. 
SMGs are highly obscured by dust, and the resulting thermal dust emission dominates the bolometric luminosity. 
The energy source of mm/submm emission is primarily from intense star formation activity, with star-formation rates (SFRs) of 10$^2$--10$^3$~$M_{\odot}$~yr$^{-1}$. 
The heavy dust obscuration in SMGs makes it difficult to understand their optical/near-infrared (NIR) properties. 
In addition, the coarse angular resolution of single dish telescopes ($\gsim$$15''$) prevents from identifying optical/NIR counterparts. 
One of the most successful ways to identify counterparts is to obtain high resolution, deep radio imaging \citep[e.g.,][]{ivis98, ivis02}.
Deep radio observations with interferometers reveal robust radio counterparts of $\sim$50\%--80\% of SMGs \citep[e.g.,][]{ivis05, ivis07}. 
A problem in radio identification is the rapid dimming of radio flux of galaxies with increasing redshift. 
The most accurate means of achieving high precision astrometry of SMGs is to use interferometers at mm/submm \citep[e.g.,][]{iono06, wang07, youn07, youn08, youn09, hats10}, although this approach needs much time to detect an object compared to the radio imaging.
The advent of the Atacama Large Millimeter/submillimeter Array (ALMA) has changed this situation thanks to its high sensitivity and high angular resolution.

Optical/NIR follow-up observations and spectral energy distribution (SED) model fits have shown that SMGs have substantial stellar masses of $\sim$$10^{11}$--$10^{12}$~$M_{\odot}$ \citep[e.g.,][]{smai04, bory05, dye08, mich10, hain11}. 
It is thought that SMGs are progenitors of massive elliptical galaxies in the present-day universe observed during their formation phase \citep[e.g.,][]{lill99, smai04, simp14}. 
It is known that star-forming galaxies follow a tight correlation between stellar mass and SFR \citep[star-forming main sequence; e.g.,][]{noes07}, and SMGs are found to be located above the main sequence or at the massive end of the main sequence \citep[e.g.,][]{dadd07, dacu15}. 
These suggest that previous mm/submm surveys trace only a small fraction of star-forming galaxies. 
This is inferable by the fact that SMGs detected in previous blank-field surveys contribute the extragalactic background light, which is the integral of unresolved emission from extragalactic sources, by only $\sim$20\%--40\% at 850~$\mu$m \citep[e.g.,][]{barg99, bory03, copp06} and $\sim$10\%--20\% at 1~mm \citep[e.g.,][]{grev04, scot08, scot10, hats11}. 
In order to understand the cosmic star-formation history and galaxy evolution, it is necessary to study fainter mm/submm sources ($S_{\rm 1 mm} \lsim 1$ mJy), which connect `classical' SMGs and `normal' star-forming galaxies. 
However, since the previous surveys of SMGs were conducted with single-dish telescopes, it has been very hard to detect fainter sources because of the limited sensitivity and the source confusion except for rare samples around gravitational lensing clusters \citep[e.g.,][]{chen14}, and it is still unclear the optical/NIR properties of fainter mm/submm sources.
ALMA enables us to detect fainter sources with the flux densities about an order of magnitude fainter than those detected in previous single-dish surveys, allowing us to study the properties of faint mm/submm sources \citep[e.g.,][]{hats13, ono14, fuji15}. 
In this paper, we present the optical--IR properties of ALMA-detected faint sources ($S_{\rm 1.3 mm} = 0.2$--1.0~mJy). 
The arrangement of this paper is as follows. 
Section~\ref{sec:data} outlines the data we used. 
Section~\ref{sec:id} describes the method of counterpart identification. 
In Section~\ref{sec:discussion}, we discuss the optical--IR properties of the ALMA-detected sources. 
A summary is presented in Section~\ref{sec:summary}. 
Throughout the paper, we adopt a cosmology with $H_0=70$ km s$^{-1}$ Mpc$^{-1}$, $\Omega_{\rm{M}}=0.3$, and $\Omega_{\Lambda}=0.7$. 
All magnitudes are in the AB system \citep{oke83}.

\section{Data}\label{sec:data}

We conducted ALMA band 6 observations toward 20 star-forming galaxies at $z \sim 1.4$ in the Subaru/{\sl XMM-Newton} Deep Survey (SXDS) field \citep{furu08}. 
The targets were extracted from a stellar mass limit ($>$$10^{9.5}$~$M_{\odot}$) sample whose redshifts and H$\alpha$ SFR were obtained by near-infrared (NIR) spectroscopy with the Fibre Multi-Object Spectro-graph \citep[FMOS;][]{kimu10} on the Subaru telescope \citep[][]{yabe12, yabe14}. 
The ALMA observations were carried out in August 2012 with 23--25 antennas during the cycle~0 session. 
The correlator was used in the frequency domain mode with a bandwidth of 1875 MHz (488.28 kHz $\times$ 3840 channels). 
We obtained 20 pointings centered on the 20 targets, each with on-source observing time of 8--15 minutes. 
The full width at half maximum (FWHM) of the primary beam is $\sim$$26''$.

The data were reduced with the Common Astronomy Software Applications \citep[CASA;][]{mcmu07} package in a standard manner. 
We found that the coordinates of a phase calibrator were wrong (by $\sim$$0.3''$) in the originally delivered data, which causes positional offsets of sources detected in science maps. 
We re-calibrated the data by modifying the phase center of the visibilities of the phase calibrator manually to obtain the correct coordinates in the final science maps. 
We used the 2012 models of Solar System Object for flux calibrations instead of the 2010 models used in \cite{hats13}, which makes the flux density of the maps at most 15\% smaller. 
The maps were processed with the {\verb CLEAN } algorithm with the natural weighting, which gives the final synthesized beamsize of $\sim$0\farcs6--1\farcs3.
The continuum images of the 20 fields are created with the rms noise level of 0.04--0.10 mJy~beam$^{-1}$.

We used the area within the primary beam of the images for source detection. 
Source extraction was conducted on the 20 continuum images (before primary beam correction) where all the sources with a peak signal-to-noise ratio (SN) above 3.5 were {\verb CLEAN }ed. 
The probability of contamination by spurious sources ($C_{\rm spurious}$) is estimated by counting the negative peaks in each map as a function of SN and averaged over the 20 images \citep{hats13}.
In this paper, we adopt the detection threshold of SN $\ge 4.0$, where $C_{\rm spurious}$ is less than 0.5.
We detected 8 sources at SN $\ge 4.0$, of which three sources are the original targets of ALMA observations and five sources are serendipitously-detected sources. 
The source list is presented in Table~\ref{tab:source}
\footnote[1]{Two sources (AS2 and AS4) are also detected  by \cite{fuji15}.}. 
The peak flux density of the continuum sources corrected for the primary beam attenuation is $S_{\rm 1.3\ mm} = 0.17$--1.0 mJy (4.0--13$\sigma$). 
Note that the source sizes in the images (without deconvolution) are not significantly smaller than the synthesized beamsize. 
In this study, we adopt the continuum flux density measured without excluding channels where CO emission lines present to be fairly compared with past and future studies in the continuum where the contamination is unknown.

\begin{table*}
\begin{center}
\caption{ALMA-detected Sources \label{tab:source}}
\begin{tabular}{cccccccc}
\tableline\tableline
Name & ID & R.A.    & Decl.   & $S_{\rm 1.3mm}^{\rm a}$ & SN & $C_{\rm spurious}^{\rm b}$ & Note \\
     &    & (J2000) & (J2000) & (mJy) & & & \\
\tableline
ALMA\_SXDS1\_13015\_1 & AS1 & 02 17 13.63 & -05 09 40.0 & $1.00 \pm 0.08$ &13   & 0.0 & original target \\
ALMA\_SXDS1\_31189\_1 & AS2 & 02 17 13.33 & -05 04 02.3 & $0.30 \pm 0.06$ & 4.8 & 0.0 & serendipitous   \\
ALMA\_SXDS1\_31189\_2 & AS3 & 02 17 13.82 & -05 04 18.7 & $0.35 \pm 0.08$ & 4.3 & 0.1 & serendipitous   \\
ALMA\_SXDS1\_59863\_1 & AS4 & 02 17 46.27 & -04 54 39.8 & $0.53 \pm 0.09$ & 6.0 & 0.0 & serendipitous   \\
ALMA\_SXDS1\_59863\_2 & AS5 & 02 17 45.89 & -04 54 37.4 & $0.30 \pm 0.07$ & 4.0 & 0.4 & original target \\
ALMA\_SXDS1\_79307\_1 & AS6 & 02 17 06.02 & -04 51 37.5 & $0.36 \pm 0.09$ & 4.1 & 0.3 & serendipitous   \\
ALMA\_SXDS3\_110465\_1& AS7 & 02 18 21.27 & -05 19 06.9 & $0.39 \pm 0.08$ & 4.8 & 0.0 & serendipitous   \\
ALMA\_SXDS5\_28019\_1 & AS8 & 02 16 08.53 & -05 06 15.8 & $0.17 \pm 0.04$ & 4.4 & 0.1 & original target \\
\tableline
\end{tabular}
\tablecomments{
$^{\rm a}$ 1.3~mm flux density corrected for primary beam attenuation. 
$^{\rm b}$ Probability of contamination by spurious sources. 
}
\end{center}
\end{table*}

\begin{figure}
\begin{center}
\includegraphics[width=\linewidth]{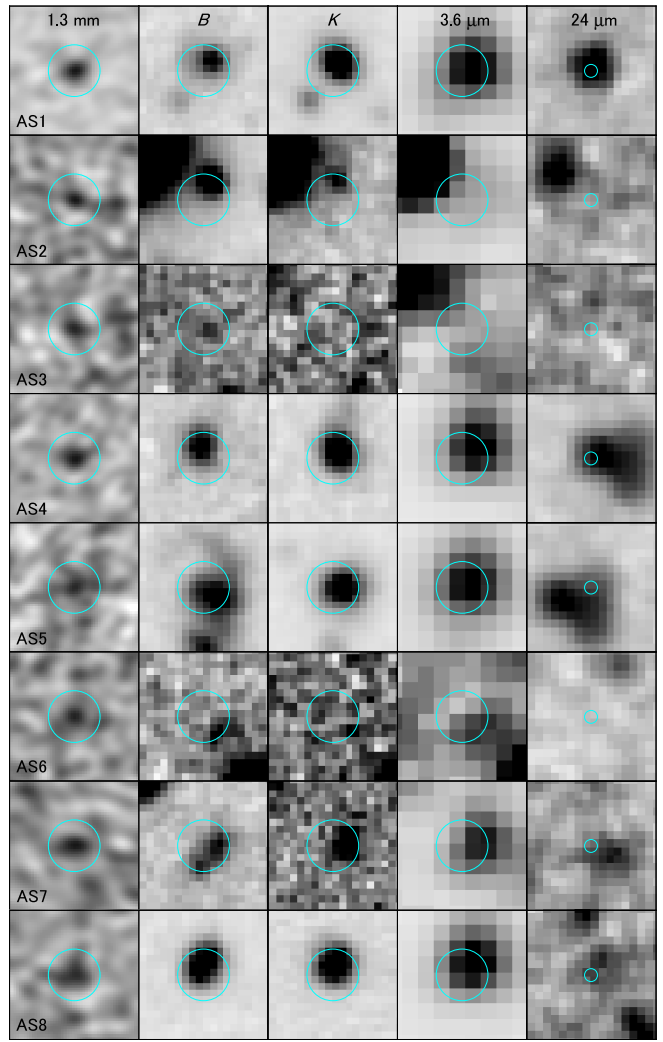}
\caption{
Multi-wavelength images of the ALMA sources. 
Panels from left to right are ALMA 1.3~mm, Subaru/Suprime-Cam $B$, UKIRT/WFCAM $K_s$, {\sl Spitzer}/IRAC 3.6~$\mu$m, and MIPS 24~$\mu$m. 
The size of each panel is $5'' \times 5''$ for the first four panels, and $20'' \times 20''$ for the last panel. 
Circles with 1$''$ radius centered on the ALMA sources are presented in each panel. 
\label{fig:stamp}}
\end{center}
\end{figure}

\begin{figure}
\begin{center}
\includegraphics[width=\linewidth]{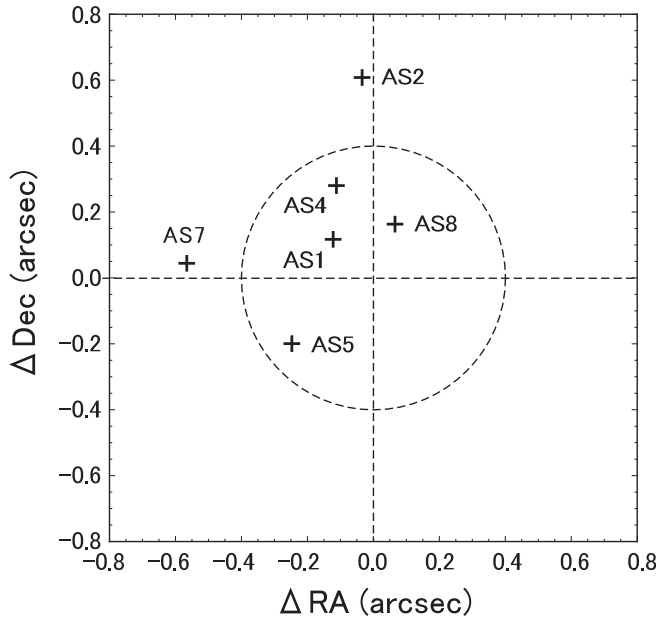}
\end{center}
\caption{
Positional offsets of the $K$-band counterpart candidates from the ALMA sources. 
The dashed circle represents the offset of 0\farcs4, which is the maximum expected positional offset between the ALMA sources and $K$-band counterparts. 
\label{fig:offset}}
\end{figure}

\section{Counterpart Identification}\label{sec:id}
Optical/IR counterparts for the ALMA sources are searched in the $K$-band-selected source catalog of \cite{yabe12, yabe14}. 
The limiting magnitude of the catalog is $K_s=24.6$~mag (2\farcs0 aperture, 5$\sigma$), and the WCS accuracy of the source coordinates is $\sim$0\farcs2--0\farcs3 (rms). 
The synthesized beam size of the ALMA observations is $\sim$0\farcs6--1\farcs3 (major axis FWHM), and the expected astrometric accuracy of the ALMA sources is $\sim$0\farcs06--0\farcs3 ($\sim$FWHM/SN). 
Therefore, the expected positional accuracy between $K$-band and ALMA coordinates is $\sim$0\farcs3--0\farcs4 as the square-root of sum of squares of both accuracies. 
\cite{hodg13} compared ALMA-detected SMGs with VLA 1.4~GHz counterparts and found that the positions are accurate to within 0\farcs2--0\farcs3. 
The average positional offsets between ALMA-detected SMGs and {\sl HST} counterparts are found to be 0\farcs3--0\farcs4 \citep{wikl14, chen15}. 
The positions of mm/sumbm emission and $K$-band emission, which typically trace dust-obscured and unobscured part, respectively, are not necessarily coincide in a galaxy \citep[e.g.,][]{iono06, chen15, hodg15, hats15}. 
Figure~\ref{fig:stamp} shows the 1.3~mm image for the 8 ALMA sources together with multi-wavelength images of 
Subaru/Suprime-Cam $B$ \citep{furu08}, 
UKIRT/Wide Field Camera (WFCAM) $K_s$ \citep{lawr07}, 
{\sl Spitzer}/Infrared Array Camera (IRAC) 3.6~$\mu$m, and MIPS 24~$\mu$m (Dunlop et al. in preparation). 
We found $K$-band counterpart candidates for 6 out of the 8 ALMA sources within a radius of $1''$. 
AS3 and AS6 have faint emission in the $B$-band image, but they are not detected in $K$-band. 
The positional offset between the ALMA sources and their $K$-band counterpart candidates is shown in Figure~\ref{fig:offset}. 
AS2 and AS7 have a larger offset (0\farcs5--0\farcs6) than the expected positional accuracy of $\sim$0\farcs3--0\farcs4, and we regard them as unidentified. 
The average offset for the remaining four ALMA sources is ($\Delta$RA, $\Delta$Dec) $= (-0\farcs10 \pm 0\farcs13, +0\farcs09 \pm 0\farcs20)$, which is within the expected positional accuracy between $K$-band and ALMA coordinates.

We identified $K$-band counterparts for four out of the 8 ALMA sources.
If we focus on the serendipitous sources, counterparts are found for 1 out of 5 sources. 
One possibility for not having $K$-band counterparts is that the ALMA sources are spurious. 
The sum of $C_{\rm spurious}$ of the 8 ALMA sources is $\sim$1 (Table~\ref{tab:source}), suggesting that at least one source is spurious. 
Another possibility is that they are obscured by dust or at higher redshift, which could make the optical/NIR emission fainter than the limiting magnitude of our $K$-band source catalog \citep{chen14}. 
The faint $B$-band emission in A3 and A6 may be from less obscured regions within the sources. 
While we expect emission at mid-IR if they have dust-obscured star-forming regions, they are not detected with {\sl Spitzer}, suggesting that they may be at higher redshift if not spurious.

\section{Discussion}\label{sec:discussion}

\begin{figure*}
\begin{center}
\includegraphics[width=\linewidth]{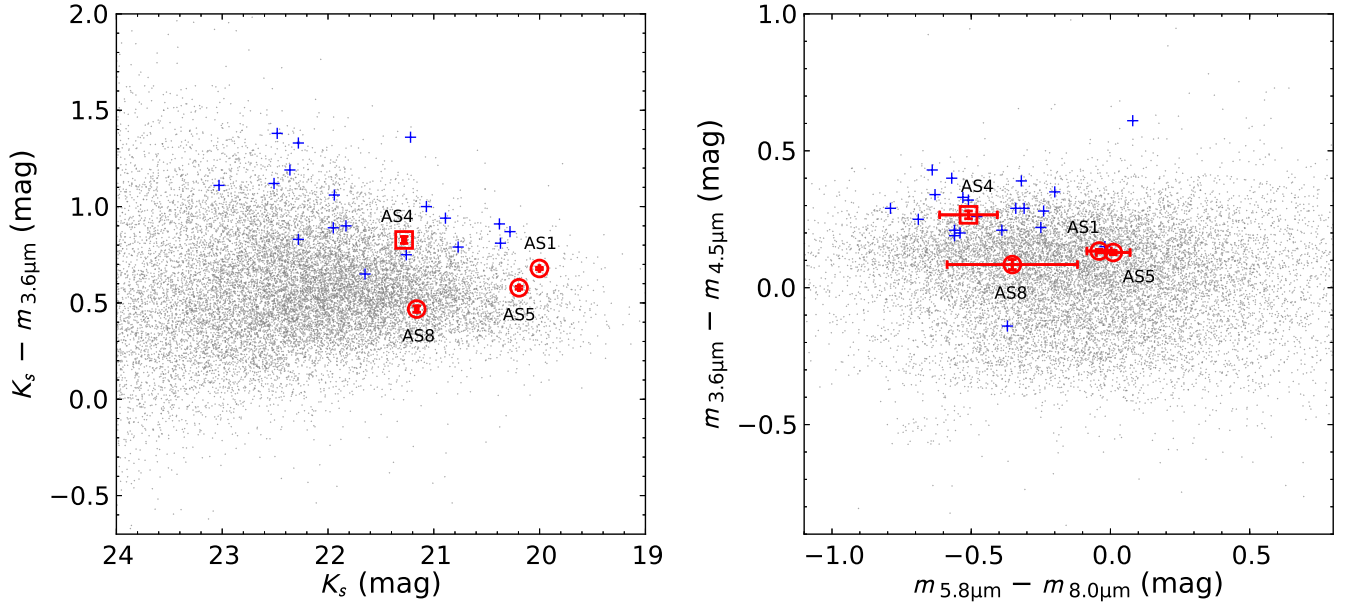}
\end{center}
\caption{
$K$-band and IRAC color-magnitude diagram ($K_s$ versus $K_s-3.6~\mu$m) (left) and IRAC color-color diagram ($m_{3.6 \rm \mu m} - m_{4.5 \rm \mu m}$ versus $m_{5.8 \rm \mu m} - m_{8.0 \rm \mu m}$) (right). 
The ALMA sources identified with the original FMOS targets and the ALMA serendipitous source are presented as circles and squares, respectively. 
For comparison, we plot ALMA-identified SMGs of \cite{simp14} (crosses) and $K$-band sources in the catalog of \cite{yabe14} (dots) at a spectroscopic (if available) or photometric redshift range of $1.0 < z < 2.0$. 
\label{fig:K-irac}}
\end{figure*}

\subsection{Optical--IR Properties}\label{sec:properties}
For the ALMA sources with $K$-band counterparts, we investigate the optical--IR properties. 
Fig.~\ref{fig:K-irac} shows a $K$-band and IRAC color-magnitude diagram ($K_s$ versus $K_s-3.6~\mu$m) (left) and a IRAC color-color diagram ($m_{3.6 \rm \mu m} - m_{4.5 \rm \mu m}$ versus $m_{5.8 \rm \mu m} - m_{8.0 \rm \mu m}$) (right) for the ALMA sources. 
For comparison, $K$-band sources at $1.0 < z < 2.0$ in the catalog of \cite{yabe14} are plotted. 
We also show SMGs identified in the ALMA follow-up observations of the LABOCA Extended Chandra Deep Field South surveys (ALESS). 
The ALESS SMGs in the same redshift range ($1.0 < z < 2.0$) are taken from \cite{simp14}. 
The plots show that our ALMA sources are bluer in $K_s - m_{3.6 \rm \mu m}$ and $m_{3.6 \rm \mu m} - m_{4.5 \rm \mu m}$ compared to the ALESS SMGs.

\begin{figure}
\begin{center}
\includegraphics[width=\linewidth]{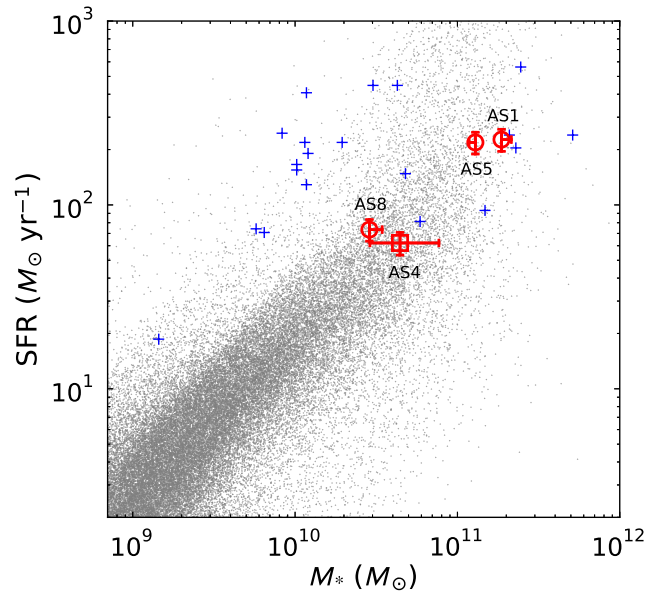}
\end{center}
\caption{
Comparison of stellar mass and SFR for the ALMA sources, ALMA-identified SMGs \citep{dacu15}, and the $K$-band sources. 
Symbols are the same as in Fig.~\ref{fig:K-irac}. 
The ALMA-identified SMGs and $K$-band sources are at a redshift range of $1.0 < z < 2.0$. 
\label{fig:stellar-SFR}}
\end{figure}

We estimated physical quantities for our ALMA sources. 
The photometric redshift and the stellar mass are derived from SED fits to far-UV--mid-IR data of Galaxy Evolution Explorer (GALEX) far- and near-UV, CFHT/MegaCam $u$, Subaru/Suprime-Cam $B$, $V$, $R_C$, $i'$, and $z$, UKIRT/WFCAM $J$, $H$ and $K_s$, {\sl Spitzer}/IRAC 3.6~$\mu$m, 4.5~$\mu$m, 5.8~$\mu$m, and 8.0~$\mu$m. 
The Salpeter IMF \citep{salp55} with a mass range of 0.1--100~$M_{\odot}$ was assumed. 
The color excess was derived from the rest-frame UV color, and the SFR was derived from the rest-frame UV luminosity density corrected for extinction by using the color excess \citep[see][for details]{yabe12, yabe14}. 
The quantities are summarized in Table~\ref{tab:properties}. 
The standard deviation of the difference between photometric and spectroscopic redshifts in the sample of \cite{yabe14} is $\sigma \sim 0.05$. 
We adopt $\Delta z = 0.05$ for the error of photometric redshifts. 
We note that the ALMA observing fields were centered on $z \sim 1.4$ star-forming galaxies, which could be biased regions. 
To eliminate such a bias completely, it is essential to observe `purely' blank fields. 
In Figure~\ref{fig:stellar-SFR}, we compared the stellar mass and SFR of our ALMA with those of $K$-band sources and the ALESS SMGs at $1.0 < z < 2.0$. 
While more than half of the ALESS SMGs are above the main sequence \citep{dacu15}, the ALMA serendipitous source is located in the main sequence. 
This suggests that ALMA-detected faint sources are likely to have properties similar to normal star-forming galaxies \citep{hats13, ono14, fuji15}.

\begin{table*}
\begin{center}
\caption{Physical properties of ALMA-detected sources with $K$-band counterpart candidates \label{tab:properties}}
\begin{tabular}{lcccccccc}
\tableline\tableline
ID & offset$^{\rm a}$ & $K_s$ & $z_{\rm spec}$ & $z_{\rm phot}$ & $E(B-V)$ & $M_*$ & SFR(UV SED)\\
   & (arcsec) & (mag) &              &                & (mag)    & ($M_{\odot}$) & ($M_{\odot}$~yr$^{-1}$) \\
\tableline
AS1          & 0.17 & $20.00 \pm 0.00$ & 1.451 & $1.42 \pm 0.05$ & $0.50^{+0.00}_{-0.05}$ & $1.87^{+0.28}_{-0.00} \times 10^{11}$ & $227 \pm 31$ \\
AS4$^{\rm b}$& 0.30 & $21.28 \pm 0.02$ & $-$   & $1.53 \pm 0.05$ & $0.60^{+0.05}_{-0.10}$ & $4.43^{+3.26}_{-1.54} \times 10^{10}$ & $ 62 \pm  9$ \\
AS5          & 0.32 & $20.19 \pm 0.01$ & 1.448 & $1.43 \pm 0.05$ & $0.45^{+0.05}_{-0.00}$ & $1.29^{+0.00}_{-0.11} \times 10^{11}$ & $219 \pm 30$ \\
AS8          & 0.18 & $21.16 \pm 0.01$ & 1.348 & $1.32 \pm 0.05$ & $0.35^{+0.05}_{-0.00}$ & $2.87^{+0.57}_{-0.00} \times 10^{10}$ & $ 73 \pm 10$ \\
\tableline
\end{tabular}
\tablecomments{
$^{\rm a}$ Positional offset between ALMA source and $K$-band counterpart candidate. 
$^{\rm b}$ Serendipitous source.
}
\end{center}
\end{table*}

\begin{figure*}
\begin{center}
\includegraphics[width=\linewidth]{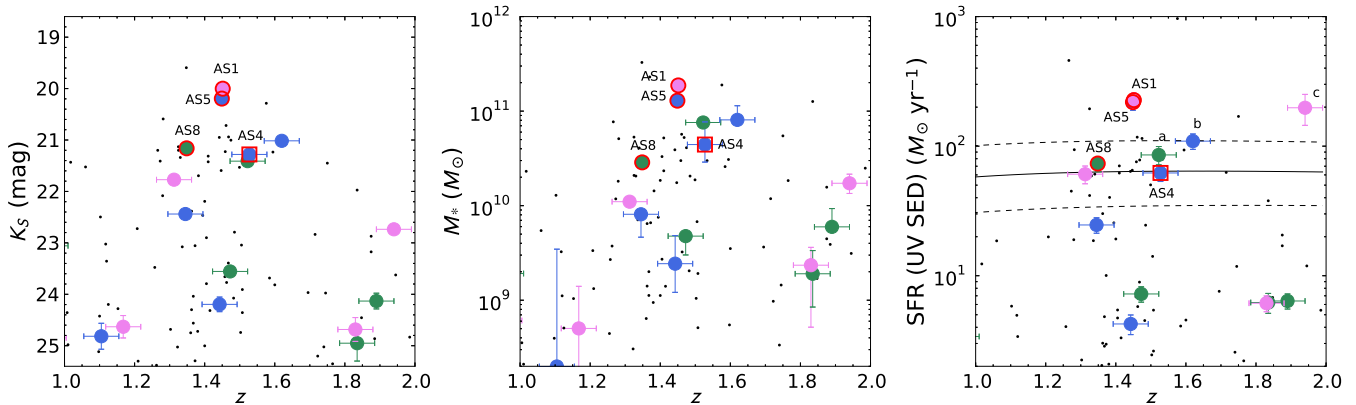}
\end{center}
\caption{
$K_s$ magnitude, stellar mass, and SFR of $K$-band sources within the field of ALMA observations as a function of redshift. 
The ALMA sources identified with the original FMOS targets and the ALMA serendipitous source are presented as red circles and squares, respectively. 
$K$-band sources in the field of views where ALMA sources are detected are shown as large circles; different colors represent different fields. 
$K$-band sources in the field of views where no ALMA sources are detected are shown as dots. 
The redshifts are spectroscopic (if available) or photometric. 
The errors on photometric redshifts ($\Delta z = 0.05$) are presented in the horizontal axis. 
In the right panel, we present SFR as a function of redshift expected for a source with $S_{\rm 1.3mm} = 0.2$~mJy, $T_{\rm dust} = 35$ ($\pm 5$~K), and $\beta = 1.5$ as a solid (dashed) line. 
\label{fig:per_field}}
\end{figure*}

\subsection{ALMA-detected and undetected sources}\label{sec:detected}

To see whether there is a common characteristic in sources that are detected with ALMA, we compared the properties of $K$-band sources detected and undetected with ALMA within the field of views of ALMA observations. 
In Figure~\ref{fig:per_field}, we plot $K$-band magnitude, stellar mass, and SFR as a function of redshift. 
The plots show that the ALMA-detected sources tend to be brighter, more massive, and more actively star-forming galaxies among the $K$-band sources. 
We show the expected SFR for a source with $S_{\rm 1.3mm} = 0.2$~mJy (the faintest flux density of our ALMA sources) as a function of redshift in the right panel of Figure~\ref{fig:per_field}.
We estimate the SFR by using the SFR--far-IR (FIR) luminosity ($L_{\rm FIR}$) relation of \cite{kenn98}. 
We derived FIR luminosity from $L_{\rm FIR} = 4\pi M_{\rm dust} \int_0^{\infty} \kappa_d(\nu_{\rm{rest}})B(\nu_{\rm{rest}}, T_{\rm dust}) d\nu$ and dust mass from $M_{\rm dust}=S_{\rm{obs}}D_L^2/[(1+z)\kappa_d(\nu_{\rm{rest}})B(\nu_{\rm{rest}}, T_{\rm dust})]$ \citep{debr03}, 
where $\kappa_d(\nu_{\rm{rest}})$ is the dust mass absorption coefficient, $\nu_{\rm{rest}}$ is the rest-frame frequency, $T_{\rm dust}$ is the dust temperature, $B(\nu_{\rm{rest}}, T_{\rm dust})$ is the Planck function, $S_{\rm{obs}}$ is the observed flux density, and $D_L$ is the luminosity distance. 
We assume that the absorption coefficient varies as $\kappa_d \propto \nu^{\beta}$ and the emissivity index lies between 1 and 2 \citep[e.g,][]{hild83}. 
We adopt $\kappa_d(125\ \mu m) = 2.64 \pm 0.29$~m$^2$~kg$^{-1}$ \citep{dunn03}, $\beta = 1.5$, and $T_{\rm dust} = 30$--40~K, typical value for $z \sim 0$--2 star-forming galaxies \citep[e.g.,][]{elba11, syme13}. 
The solid and dashed lines indicate that sources with $S_{\rm 1.3mm} \ge 0.2$~mJy above the lines can be detected with the ALMA observations based on the assumption that the SFRs derived from UV continuum and dust emission are consistent. 
There are $K$-band sources within or above the lines which are not detected with ALMA (labeled as a, b, and c in Figure~\ref{fig:per_field} (right)). 
Source a is located in the same field as AS8. 
While AS8 is located at the field center, source a is $\sim$$10''$ away from the field center, where the sensitivity is $\sim$35\% lower than that at the field center, suggesting that the lower sensitivity missed source a. 
Source b is seen in the ALMA image, but the SN of 3.9 is lower than the threshold of 4.0 adopted in this study. 
Around this source, two ALMA sources (AS4 and AS5) are detected within a narrow redshift range, suggesting that star-forming activity is enhanced by the interaction (see Section~\ref{sec:overdense}). 
Source c is located $\sim$$3''$ away from the center of the same field as AS1, where the rms noise level (0.08 mJy~beam$^{-1}$) is higher among the ALMA observing fields. 
In spite of the higher SFR, source c was not detected with ALMA, suggesting the SFR has uncertainty. 
We also investigated the properties of $K$-band sources above the the expected SFR line in Fig.~\ref{fig:per_field} in the field of views where no ALMA sources are detected, and found that the non-detection with ALMA can be explained by lower sensitivity, or lower SFRs than those of ALMA sources with the uncertainty of SFRs estimated both from UV continuum and dust emission.

\begin{table*}
\begin{center}
\caption{Physical properties of $K$-band sources in the overdense region \label{tab:overdense}}
\begin{tabular}{cccccccc}
\tableline\tableline
$K$-band ID  & $S_{\rm 1.3mm}$& $C_{\rm spurious}$& $K_s$ & $z_{\rm phot}$ & $E(B-V)$ & $M_*$ & SFR(UV SED)\\
             & (mJy)          &                   & (mag) &                & (mag)    & ($M_{\odot}$) & ($M_{\odot}$~yr$^{-1}$)\\
\tableline
SXDS1\_59617 & $0.33 \pm 0.09$ & 0.9 & $21.02 \pm 0.02$ & $1.62 \pm 0.05$ & $0.35^{+0.00}_{-0.05}$ & $8.08^{+3.32}_{-0.74} \times 10^{10}$ & $109 \pm 15$ \\
SXDS1\_60181 & $<$$0.23^{*}$   & $-$ & $22.44 \pm 0.04$ & $1.35 \pm 0.05$ & $0.40^{+0.05}_{-0.00}$ & $8.11^{+0.79}_{-3.47} \times 10^{9 }$ & $ 25 \pm  3$ \\
SXDS1\_60394 & $<$$0.23^{*}$   & $-$ & $23.55 \pm 0.09$ & $2.37 \pm 0.05$ & $0.55^{+0.05}_{-0.05}$ & $7.05^{+6.85}_{-2.34} \times 10^{9 }$ & $134 \pm 30$ \\
\tableline
\end{tabular}
\tablecomments{
$^{*}$ 3$\sigma$ upper limit. 
}
\end{center}
\end{table*}

\subsection{Overdense around ALMA sources}\label{sec:overdense}

\begin{figure}
\begin{center}
\includegraphics[width=\linewidth]{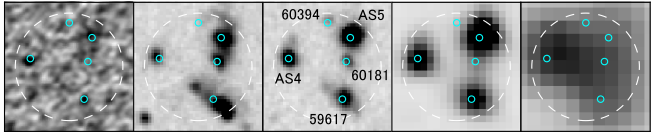}
\end{center}
\caption{
Multi-wavelength images around the multiple ALMA sources. 
From left to right: ALMA 1.3~mm, Subaru/Suprime-Cam $B$, UKIRT/WFCAM $K_s$, {\sl Spitzer}/IRAC 3.6~$\mu$m, and MIPS 24~$\mu$m. 
The dashed circle shows a region within a radius of $5''$ (42 kpc at $z = 1.45$). 
The small circles represent the positions of $K$-band sources. 
\label{fig:overdense}}
\end{figure}

We found a region where multiple ALMA sources and $K$-band sources reside in a small projected area (Figure~\ref{fig:overdense}, Table~\ref{tab:overdense}).
Five $K$-band sources are located within a radius of $5''$ (42~kpc if we assume $z = 1.45$), of which two sources are the ALMA sources (AS4 and AS5). 
SXDS1\_59617 is also presented in the ALMA 1.3 mm image with SN $= 3.9$. 
Although the probability of contamination by spurious sources for SXDS1\_59617 is 0.9 due to the low SN, the detection in other bands ($B$ or {\sl Spitzer}) suggests that the source is real. 
AS5 is the original target for ALMA observations, whose spectroscopic redshift is $z_{\rm spec} = 1.448$. 
The spectroscopic or photometric redshifts of four sources (AS4, AS5, SXDS1\_59617 and SXDS1\_60181) are $\sim$1.3--1.6, suggesting the presence of an overdense region by considering the uncertainty of photometric redshifts, although the apparent condensation of the galaxies may be chance coincidence. 
SXDS1\_60394 with $z_{\rm phot} = 2.37$ may be a background source. 
The expected number of $K$-band sources with $K_s \le 23.55$~mag (the magnitude of the faintest source in this region) or with $K_s \le 22.44$~mag (the magnitude of the faintest source excluding SXDS1\_60394) which fall in an area with a radius of $5''$ is 0.7 and 0.3, respectively. 
In a similar way, the expected number of submm sources with $S_{\rm 1.3mm} \ge 0.2$~mJy is $\sim$0.1--0.2 \citep[based on the 1.3~mm number counts of][]{hats13}. 
These suggest that this region has an excess of $K$-band sources and submm sources. 
AS5 appears to have a hint of tidal feature in the $B$-band image, suggesting an interaction. 
This region is also detected in {\sl Spitzer} 24~$\mu$m and with {\sl Herschel} (though as a single source due to its limited angular resolution), suggesting that the star-forming activity is increased. 
It is possible that these sources are a pre-merging system, and we may be witnessing the early phase of formation of a massive elliptical galaxy. 
Assuming that at least two sources including AS5 merge into a galaxy, the stellar mass would be $10^{11}$--$10^{12}$~$M_{\odot}$. 
It has been difficult to find such pre-merging systems in previous mm/submm observations due to the limited sensitivity and angular resolution before ALMA.

\section{Summary}\label{sec:summary}

We studied the optical--IR properties of 8 ALMA 1.3~mm faint sources detected in our deep observations in SXDS. 
We searched for $K$-band counterparts and found counterpart candidates for four ALMA sources within a radius of $0\farcs4$. 
The sum of probability of contamination by spurious sources for the 8 ALMA sources is $\sim$1, and it is likely that at least one them is spurious. 
Possible reasons for no $K$-band counterparts are
the ALMA detection is spurious, 
the optical/NIR color is bluer and faint in $K_s$, 
a larger offset between 1.3~mm emission and $K_s$-band emission, 
obscured by dust, 
and at higher redshift. 
While more than half of the ALMA-identified SMGs are above the main sequence, the ALMA serendipitous source is located in the main-sequence, suggesting that the ALMA-detected faint sources have properties similar to `normal' star-forming galaxies. 
Comparison between $K$-band sources detected and undetected with ALMA within the field of views suggests that ALMA-detected sources tend to be brighter in $K$-band, more massive, and more actively star-forming galaxies.

We found a region where multiple ALMA sources and $K$-band sources reside within a radius of $5''$ and in a photometric redshift range of $\sim$1.3--1.6. 
This may be a pre-merging system and the early phase of formation of a massive elliptical galaxy.

\acknowledgments

We would like to acknowledge staffs at the ALMA Regional Center for their help in data reduction. 
BH is supported by JSPS KAKENHI Grant Number 15K17616. 
KO is supported by the Grant-in-Aid for Scientific Research (C)(24540230) from the Japan Society for the Promotion of Science. 
RM is supported by the Grant-in-Aid for JSPS Fellows. 
This paper makes use of the following ALMA data: ADS/JAO.ALMA\#2011.0.00648.S. 
ALMA is a partnership of ESO (representing its member states), NSF (USA) and NINS (Japan), together with NRC (Canada), NSC and ASIAA (Taiwan), and KASI (Republic of Korea), in cooperation with the Republic of Chile. The Joint ALMA Observatory is operated by ESO, AUI/NRAO and NAOJ.

{\it Facilities:} \facility{ALMA}.


\appendix

\begin{table}
\rotatebox{90}{
\begin{minipage}{\textheight}
\fontsize{4pt}{0pt}\selectfont
\begin{center}
\caption{Photometry of ALMA-detected sources with $K$-band counterpart candidates \label{tab:photometry}} 
\begin{tabular}{ccccccccccccccc}
\tableline\tableline
Name & ID & $B$ & $V$ & $R$ & $i$ & $z$ & $J$ & $H$ & $K_s$ & $3.6~\mu m$ & $4.5~\mu m$ & $5.8~\mu m$ & $8.0~\mu m$ & $24~\mu m$ \\
\tableline
ALMA\_SXDS1\_13015\_1  & AS1 & $24.63 \pm 0.02$ & $24.28 \pm 0.02$ & $23.75 \pm 0.03$ & $23.16 \pm 0.01$ & $22.41 \pm 0.01$ & $21.17 \pm 0.01$ & $20.52 \pm 0.01$ & $20.00 \pm 0.00$ & $19.32 \pm 0.00$ & $19.19 \pm 0.00$ & $19.36 \pm 0.03$ & $19.40 \pm 0.03$ & $194 \pm 6$ \\
ALMA\_SXDS1\_59863\_1  & AS4 & $25.19 \pm 0.04$ & $24.68 \pm 0.03$ & $24.60 \pm 0.06$ & $24.08 \pm 0.02$ & $23.62 \pm 0.03$ & $22.39 \pm 0.03$ & $21.89 \pm 0.04$ & $21.28 \pm 0.02$ & $20.45 \pm 0.01$ & $20.19 \pm 0.01$ & $20.01 \pm 0.05$ & $20.52 \pm 0.09$ &  - \\
ALMA\_SXDS1\_59863\_2  & AS5 & $24.12 \pm 0.01$ & $23.89 \pm 0.02$ & $23.33 \pm 0.02$ & $22.84 \pm 0.01$ & $22.23 \pm 0.01$ & $21.15 \pm 0.01$ & $20.57 \pm 0.01$ & $20.19 \pm 0.01$ & $19.62 \pm 0.01$ & $19.49 \pm 0.01$ & $19.74 \pm 0.04$ & $19.73 \pm 0.04$ &  - \\
ALMA\_SXDS5\_28019\_1  & AS8 & $24.30 \pm 0.01$ & $24.08 \pm 0.02$ & $23.57 \pm 0.02$ & $23.28 \pm 0.01$ & $22.67 \pm 0.01$ & $22.00 \pm 0.02$ & $21.53 \pm 0.02$ & $21.16 \pm 0.01$ & $20.69 \pm 0.01$ & $20.61 \pm 0.01$ & $21.19 \pm 0.13$ & $21.55 \pm 0.19$ &  $41 \pm 4$ \\
\tableline\tableline
\end{tabular}
\end{center}
\end{minipage}
}
\end{table}

\begin{thebibliography}{}
\bibitem[Barger et al.(1999)]{barg99} Barger, A.~J., Cowie, L.~L., \& Sanders, D.~B.\ 1999, \apjl, 518, L5
\bibitem[Blain et al.(2002)]{blai02} Blain, A.~W., Smail, I., Ivison, R.~J., Kneib, J.-P., \& Frayer, D.~T.\ 2002, \physrep, 369, 111 
\bibitem[Borys et al.(2003)]{bory03} Borys, C., Chapman, S., Halpern, M., \& Scott, D.\ 2003, \mnras, 344, 385
\bibitem[Borys et al.(2005)]{bory05} Borys, C., Smail, I., Chapman, S.~C., et al.\ 2005, \apj, 635, 853
\bibitem[Chen et al.(2014)]{chen14} Chen, C.-C., Cowie, L.~L., Barger, A.~J., Wang, W.-H., \& Williams, J.~P.\ 2014, \apj, 789, 12 
\bibitem[Chen et al.(2015)]{chen15} Chen, C.-C., Smail, I., Swinbank, A.~M., et al.\ 2015, \apj, 799, 194
\bibitem[Coppin et al.(2006)]{copp06} Coppin, K., Chapin, E.~L., Mortier, A.~M.~J., et al.\ 2006, \mnras, 372, 1621
\bibitem[da Cunha et al.(2015)]{dacu15} da Cunha, E., Walter, F., Smail, I.~R., et al.\ 2015, \apj, 806, 110
\bibitem[Daddi et al.(2007)]{dadd07} Daddi, E., Dickinson, M., Morrison, G., et al.\ 2007, \apj, 670, 156
\bibitem[De Breuck et al.(2003)]{debr03} De Breuck, C., Neri, R., Morganti, R., et al.\ 2003, \aap, 401, 911 
\bibitem[Dunne et al.(2003)]{dunn03} Dunne, L., Eales, S.~A., \& Edmunds, M.~G.\ 2003, \mnras, 341, 589 
\bibitem[Dye et al.(2008)]{dye08} Dye, S., Eales, S.~A., Aretxaga, I., et al.\ 2008, \mnras, 386, 1107 
\bibitem[Elbaz et al.(2011)]{elba11} Elbaz, D., Dickinson, M., Hwang, H.~S., et al.\ 2011, \aap, 533, AA119 
\bibitem[Fujimoto et al.(2015)]{fuji15} Fujimoto, S., Ouchi, M., Ono, Y., et al.\ 2015, arXiv:1505.03523
\bibitem[Furusawa et al.(2008)]{furu08} Furusawa, H., Kosugi, G., Akiyama, M., et al.\ 2008, \apjs, 176, 1 
\bibitem[Greve et al.(2004)]{grev04} Greve, T.~R., Ivison, R.~J., Bertoldi, F., Stevens, J.~A., Dunlop, J.~S., Lutz, D., \& Carilli, C.~L.\ 2004, \mnras, 354, 779 
\bibitem[Hainline et al.(2011)]{hain11} Hainline, L.~J., Blain, A.~W., Smail, I., et al.\ 2011, \apj, 740, 96
\bibitem[Hatsukade et al.(2011)]{hats11} Hatsukade, B., et al.\ 2011, \mnras, 411, 102 
\bibitem[Hatsukade et al.(2010)]{hats10} Hatsukade, B., Iono, D., Akiyama, T., et al.\ 2010, \apj, 711, 974
\bibitem[Hatsukade et al.(2013)]{hats13} Hatsukade, B., Ohta, K., Seko, A., Yabe, K., \& Akiyama, M.\ 2013, \apjl, 769, L27 
\bibitem[Hatsukade et al.(2015)]{hats15} Hatsukade, B., Tamura, Y., Iono, D., et al.\ 2015, \pasj, in press (arXiv:1503.07997)
\bibitem[Hildebrand(1983)]{hild83} Hildebrand, R.~H.\ 1983, \qjras, 24, 267 
\bibitem[Hodge et al.(2013)]{hodg13} Hodge, J.~A., Karim, A., Smail, I., et al.\ 2013, \apj, 768, 91 
\bibitem[Hodge et al.(2015)]{hodg15} Hodge, J.~A., Riechers, D., Decarli, R., et al.\ 2015, \apjl, 798, LL18 
\bibitem[Hughes et al.(1997)]{hugh97} Hughes, D.~H., Dunlop, J.~S., \& Rawlings, S.\ 1997, \mnras, 289, 766 
\bibitem[Iono et al.(2006)]{iono06} Iono, D., Peck, A.~B., Pope, A., et al.\ 2006, \apjl, 640, L1
\bibitem[Ivison et al.(2002)]{ivis02} Ivison, R.~J., Greve, T.~R., Smail, I., et al.\ 2002, \mnras, 337, 1
\bibitem[Ivison et al.(2005)]{ivis05} Ivison, R.~J., Smail, I., Dunlop, J.~S., et al.\ 2005, \mnras, 364, 1025
\bibitem[Ivison et al.(2007)]{ivis07} Ivison, R.~J., Greve, T.~R., Dunlop, J.~S., et al.\ 2007, \mnras, 380, 199
\bibitem[Ivison et al.(1998)]{ivis98} Ivison, R.~J., Smail, I., Le Borgne, J.-F., et al.\ 1998, \mnras, 298, 583
\bibitem[Kennicutt(1998)]{kenn98} Kennicutt, R.~C., Jr.\ 1998, \araa, 36, 189 
\bibitem[Kimura et al.(2010)]{kimu10} Kimura, M., Maihara, T., Iwamuro, F., et al.\ 2010, \pasj, 62, 1135 
\bibitem[Lagache et al.(2005)]{laga05} Lagache, G., Puget, J.-L., \& Dole, H.\ 2005, \araa, 43, 727 
\bibitem[Lawrence et al.(2007)]{lawr07} Lawrence, A., Warren, S.~J., Almaini, O., et al.\ 2007, \mnras, 379, 1599 
\bibitem[Lilly et al.(1999)]{lill99} Lilly, S.~J., Eales, S.~A., Gear, W.~K.~P., et al.\ 1999, \apj, 518, 641
\bibitem[Le Floc'h et al.(2005)]{lefl05} Le Floc'h, E., et al.\ 2005, \apj, 632, 169
\bibitem[McMullin et al.(2007)]{mcmu07} McMullin, J.~P., Waters, B., Schiebel, D., Young, W., \& Golap, K.\ 2007, Astronomical Data Analysis Software and Systems XVI, 376, 127 
\bibitem[Micha{\l}owski et al.(2010)]{mich10} Micha{\l}owski, M., Hjorth, J., \& Watson, D.\ 2010, \aap, 514, A67 
\bibitem[Noeske et al.(2007)]{noes07} Noeske, K.~G., Weiner, B.~J., Faber, S.~M., et al.\ 2007, \apjl, 660, L43 
\bibitem[Oke \& Gunn(1983)]{oke83} Oke, J.~B., \& Gunn, J.~E.\ 1983, \apj, 266, 713 
\bibitem[Ono et al.(2014)]{ono14} Ono, Y., Ouchi, M., Kurono, Y., \& Momose, R.\ 2014, \apj, 795, 5
\bibitem[Salpeter(1955)]{salp55} Salpeter, E.~E.\ 1955, \apj, 121, 161 
\bibitem[Scott et al.(2008)]{scot08} Scott, K.~S., Austermann, J.~E., Perera, T.~A., et al.\ 2008, \mnras, 385, 2225 
\bibitem[Scott et al.(2010)]{scot10} Scott, K.~S., et al.\ 2010, \mnras, 405, 2260 
\bibitem[Simpson et al.(2014)]{simp14} Simpson, J.~M., Swinbank, A.~M., Smail, I., et al.\ 2014, \apj, 788, 125
\bibitem[Smail et al.(2004)]{smai04} Smail, I., Chapman, S.~C., Blain, A.~W., \& Ivison, R.~J.\ 2004, \apj, 616, 71 
\bibitem[Smail et al.(2014)]{smai14} Smail, I., Geach, J.~E., Swinbank, A.~M., et al.\ 2014, \apj, 782, 19
\bibitem[Symeonidis et al.(2013)]{syme13} Symeonidis, M., Vaccari, M., Berta, S., et al.\ 2013, \mnras, 431, 2317 
\bibitem[Takeuchi et al.(2005)]{take05} Takeuchi, T.~T., Buat, V., \& Burgarella, D.\ 2005, \aap, 440, L17
\bibitem[Wang et al.(2007)]{wang07} Wang, W.-H., Cowie, L.~L., van Saders, J., Barger, A.~J., \& Williams, J.~P.\ 2007, \apjl, 670, L89
\bibitem[Wiklind et al.(2014)]{wikl14} Wiklind, T., Conselice, C.~J., Dahlen, T., et al.\ 2014, \apj, 785, 111 
\bibitem[Yabe et al.(2012)]{yabe12} Yabe, K., Ohta, K., Iwamuro, F., et al.\ 2012, \pasj, 64, 60 
\bibitem[Yabe et al.(2014)]{yabe14} Yabe, K., Ohta, K., Iwamuro, F., et al.\ 2014, \mnras, 437, 3647
\bibitem[Younger et al.(2008)]{youn08} Younger, J.~D., Dunlop, J.~S., Peck, A.~B., et al.\ 2008, \mnras, 387, 707
\bibitem[Younger et al.(2007)]{youn07} Younger, J.~D., Fazio, G.~G., Huang, J.-S., et al.\ 2007, \apj, 671, 1531
\bibitem[Younger et al.(2009)]{youn09} Younger, J.~D., Fazio, G.~G., Huang, J.-S., et al.\ 2009, \apj, 704, 803
\end{thebibliography}
\end{document}